%% file: main.tex
\include{shortcuts}

\documentclass{ptapap}

\usepackage{color}
\usepackage{soul}
\definecolor{vero}{rgb}{1.0, 0.50, 0.75}

\definecolor{christi}{rgb}{0.0, 0.58, 0.71}

\author{Christiana Erba}[UD]
\author{V\'eronique Petit}[UD]
\author{Alexandre David-Uraz}[UD]
\author{Alex Fullerton}[stsci]
\affil[UD]{Department of Physics and Astronomy, University of Delaware, Newark, DE 19716, USA}
\affil[stsci]{Space Telescope Science Institute, Baltimore, MD 21218, USA}

\title{Quantitative Modeling of the UV Line Profiles of Magnetic Massive Stars}

\begin{document}

\maketitle

\begin{abstract}


Recent spectropolarimetric surveys (MiMeS, BOB) have revealed that approximately 7\% of massive stars host stable, surface dipolar magnetic fields with strengths on the order of kG. These fields channel the dense radiatively driven stellar wind into a circumstellar magnetosphere. Wind-sensitive UV spectral lines can probe the density and velocity structure of massive star magnetospheres, providing insight into wind-field interactions. To date, large-scale 
magnetohydrodynamic modeling of this phenomenon has been limited by the associated computational cost. Our analysis, using the Analytic Dynamical Magnetosphere model, solves this problem by applying a simple analytic prescription to efficiently calculate synthetic UV spectral lines. It can therefore be applied in the context of a larger parameter study to derive the wind properties for the population of known magnetic O stars. 
We also present the latest UV spectra of the magnetic O star NGC 1624-2 obtained with HST/COS, which test the limits of our models and suggest a particularly complex magnetospheric structure for this archetypal object. 

\end{abstract}


\section{Introduction}


The UV spectra of massive stars are characterized by the presence of spectral lines with P Cygni profiles, a clear signature of stellar outflows. Modeling these wind-sensitive resonance lines is the key to determining the kinematic properties of the wind and the associated mass-loss rate, a critical component of massive star evolution.

The sensitivity of line profiles to wind kinematics is particularly useful for examining the wind structure of magnetic massive stars, which host stable, nearly dipolar fields with surface field strengths typically on the order of a few kG.
Their large-scale fields influence stellar evolution by channeling outflowing material along magnetic field lines, forming dynamic magnetospheres \citep{udDoula_2002} that impact mass-loss and rotation. The resulting break in spherical symmetry in both density and flow velocity affects the shape of the UV line profiles; as a result, spherically symmetric models of the UV spectra of magnetic massive stars cannot be used to accurately derive 
wind parameters. 

The UV spectra of non-magnetic O stars generally exhibit line saturation and deep absorption troughs in some strong resonance lines (e.g. C~\textsc{iv} $\lambda\lambda$1548/50\,\AA). In contrast, UV spectra of magnetic O stars such as HD 108 \citep{Marcolino_2012}, HD 191612 \citep{Marcolino_2013}, CPD -28$^\circ$ 2561 \citep{Naze__2015}, and NGC 1624-2 \citep{DavidUraz_2019} have uncharacteristically unsaturated lines for their spectral types. Furthermore, their spectral lines are often observed to vary with rotation. This is because the magnetic pole is not generally aligned with the stellar rotation axis,  which yields a different view of the magnetosphere as the star rotates. 
To date, UV line synthesis performed using the output from magnetohydrodynamical (MHD) simulations \citep[e.g.][]{udDoula_2002} have been successful in qualitatively reproducing the behavior of the line profiles \citep{Marcolino_2013, Naze__2015}. However this method is too computationally expensive for a larger quantitative study. 

We have therefore undertaken a systematic investigation of the formation of UV resonance lines in magnetic massive stars, by pairing the parametrized Analytic Dynamical Magnetosphere (ADM) formalism \citep{Owocki_2016} with a simple radiative transfer scheme for line scattering. The synthetic line profiles we produce can be compared to observed spectra and used to both develop better diagnostics of the wind and magnetospheric parameters of magnetic massive stars and to design observational strategies to study the growing population of magnetic OB stars.


\section{Modeling Techniques}

\begin{figure}
\centering
\includegraphics[scale=0.5]{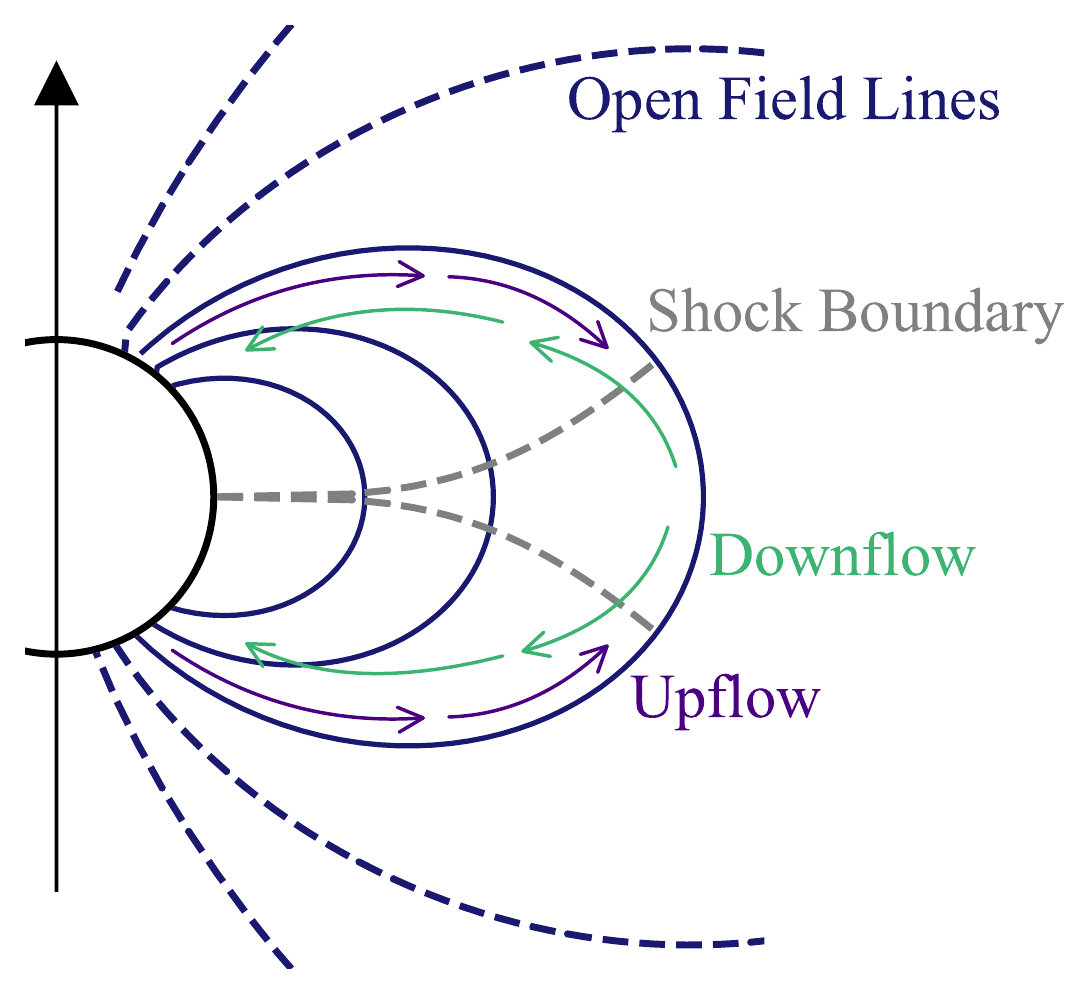}
\caption{Cartoon depicting a star with a dipolar magnetic field (the magnetic axis corresponds to the black arrow) and its magnetosphere. The two components described by the ADM model that are important to UV radiative transfer are depicted: the upflow (purple) and the downflow (green). The hot post-shock gas is located between the two shock boundaries (depicted in grey), but does not contribute to the radiative transfer calculation. The open field region is considered to only contain upflowing material. A pole-on viewing angle ($\alpha = 0^{\circ}$) places the observer along the magnetic axis; an equator-on viewing angle ($\alpha = 90^{\circ}$) places the observer at the magnetic equator.} 
\label{fig:toon}
\end{figure}



The ADM model is a physically motivated, analytic description of the time-averaged mass flow within the closed magnetic loops of a centered dipolar magnetic field, under the assumption that stellar rotation is not dynamically significant to the structure of the magnetosphere.

In this model, the mass flow within closed magnetic loops is divided into three distinct components, as illustrated in Fig. \ref{fig:toon}:

    \begin{itemize}
    \item[(i)] The wind \textit{upflow} consists of material that is radiatively driven from the stellar surface and is channeled along the field lines towards the magnetic equator;
    \item[(ii)] The \textit{hot post-shock gas} is the result of the collision of upflowing material from each magnetic hemisphere. Because of the finite cooling length of the shock-heated plasma, this region extends from a \textit{shock boundary} (see dashed grey line in Fig. \ref{fig:toon}) to the magnetic equator
    ; and
    \item[(iii)] The wind \textit{downflow} consists of the cooled post-shock material that flows back from the magnetic equator to the stellar surface under only the influence of stellar gravity.
    \end{itemize}
 
 In the model, upflow and downflow material are taken to co-exist at any point in space within the wind confinement region.
While it is not physical, this superposition of regions is justified in the time-averaged view by the short dynamical timescale of stellar winds and the quasi-independence of the field lines within the confined region. 
 To date, the ADM model has been validated with 2D MHD simulations, and has been successfully used to reproduce quantitatively the behavior of X-ray and \ha emission of magnetic OB stars \citep{Naze__2015,Owocki_2016}.


The ADM formalism has been coupled with a sophisticated radiative transfer scheme developed by \citet{Hennicker_2018}, who found a reasonable agreement between the synthetic UV profiles computed using their method, both from a MHD simulation and from an ADM magnetosphere. However in the latter case, the upflow and downflow components of the model could not be taken into account simultaneously along individual magnetic field loops. 

In our method, we compute synthetic line profiles by performing the radiative transfer along a predetermined set of rays for which the density and velocity are computed analytically at each point, treating both the upflow and downflow components (where applicable). However, we use the simplifying approximation that the source function is optically thin \citep{Owocki_1985}. We find good agreement with the profiles presented by \citet{Hennicker_2018}.



We assume that the wind material in the hot post-shock gas is ionized beyond the species present in the UV spectrum, and thus does not contribute to the formation of the lines considered in this study. 
In the open-field region (where the kinetic energy density of the wind overcomes the magnetic energy density of the field), we assume for simplicity that the wind still follows the dipole field topology, 
but we model this region by applying only the upflow component, as the material escapes the star. This approximation is adequate when magnetic confinement is strong. 


\begin{figure}
\centering
\includegraphics[scale=0.5]{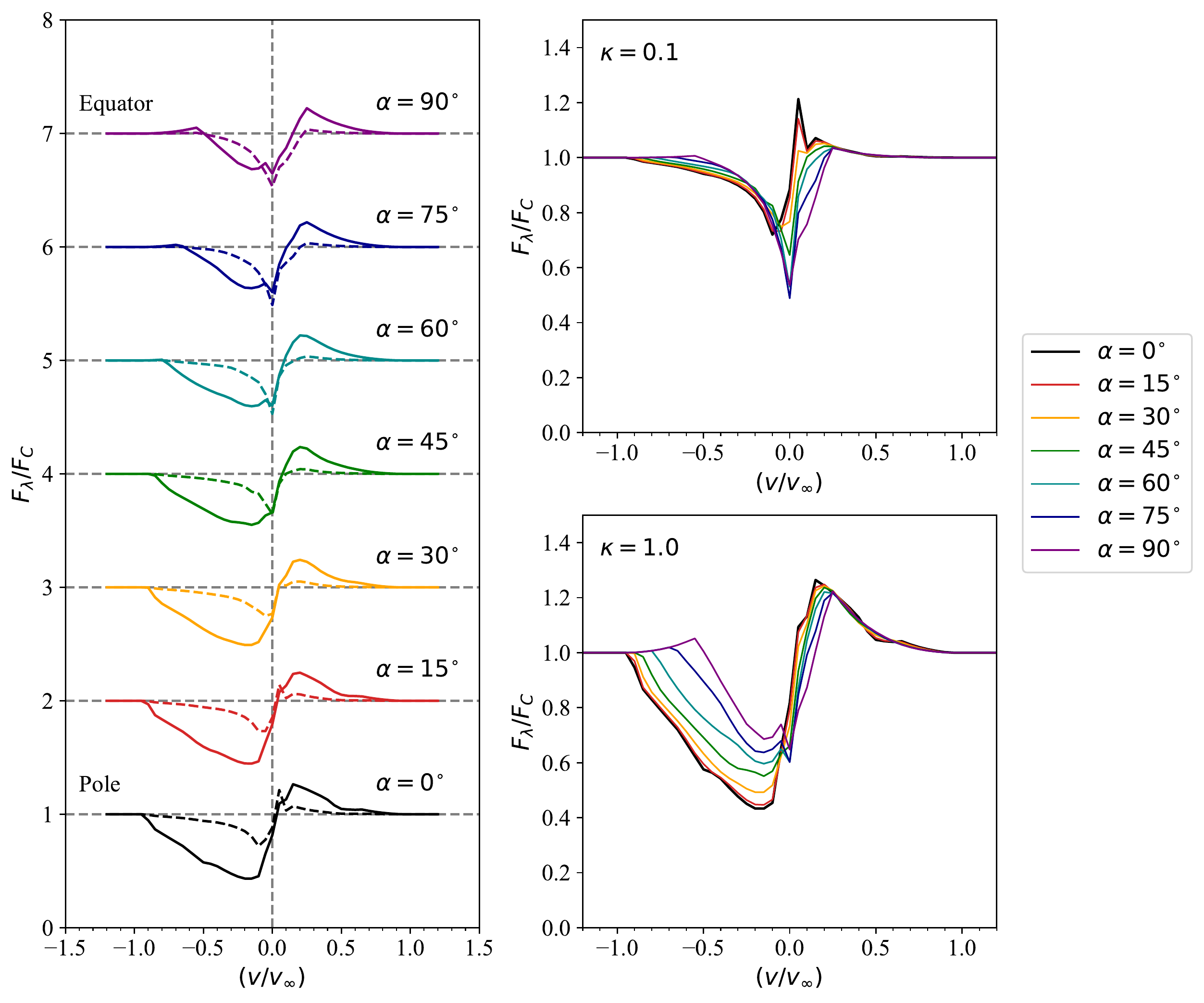}
\caption{\label{fig:kappa}Synthetic UV line profiles computed with characteristics similar to those of the magnetic O-type star HD 191612. The left panel shows, from bottom to top, the progression from a pole-on view ($\alpha=0^{\circ}$) to an equator-on view ($\alpha=90^{\circ}$), comparing the profiles for a moderately strong line ($\kappa_0=1.0$, solid curves) and a weak line ($\kappa_0=0.1$, dashed curves). The two panels on the right-hand side directly compare the viewing angle variation for the two line strengths.}

\end{figure}

We present synthetic line profiles using characteristics similar to that of the magnetic O star HD 191612 \citep{Howarth_2007,Wade_2011}. 
Figure \ref{fig:kappa} (left panel) shows the variation of the line profile from pole to equator for a weak line and for a moderately strong line (respectively $\kappa_0=0.1$ in dashed curves and $\kappa_0=1.0$ in solid curves, following the formalism of \citealt{2014A&A...568A..59S}). 
Unsurprisingly, the weak line shows considerably less overall absorption than the moderately strong line at all viewing angles. 
The weak lines also produce negligible emission when compared to the moderately strong line.

The right panel of Figure \ref{fig:kappa} reproduces the profiles in the left panel with the phase variation overplotted for the weak line (top) and the moderately strong line (bottom). 
At $\kappa_0=1.0$, the absorption from high velocity material at the blue edge of the line profile decreases as the viewing angle increases.
At $\kappa_0=0.1$, the absorption from high velocity material at the blue edge of the line profile also marginally decreases as the viewing angle increases. However at line center, the absorption is stronger when the viewing angle is small, leading to more overall absorption when the star is viewed equator-on. 
This result mirrors the modeling result from MHD simulations reported by \citet{Marcolino_2013}. 
There, the authors conclude this effect can be explained by weak lines probing the higher density, lower velocity material close to the stellar surface, a phenomenology also detailed by \citet{Naze__2015}.

Therefore, the detailed variations of spectral line profiles can be used to quantitatively diagnose the wind parameters of massive stars: since the \textit{line strength} parameter depends not only on the oscillator strength of the associated transition but also on the mass-loss rate and terminal velocity of the wind, modeling these variations for an ensemble of lines yields promising predictive power.


\section{Initial Results}



One important advantage of our new approach is that it can probe a regime that cannot be properly modeled using MHD. Indeed, for stars with very strong magnetic fields, extremely short Courant stepping times render this type of MHD calculation impractical. As a result, in
our initial analysis \citep{Erba_2017,DavidUraz_2019}, we focused on the most strongly magnetized O star observed to date, NGC 1624-2, which hosts a surface magnetic field of approximately 20 kG \citep{Wade_2012}. \citet{DavidUraz_2019} published the first UV spectra of NGC 1624-2, which showed line profiles that vary strongly with rotational phase and are significantly distinct from those produced by a spherically symmetric wind. Using the ADM prescription, the behavior of the high-velocity absorption trough was successfully modeled.

\begin{figure}
\centering
\includegraphics[scale=0.25]{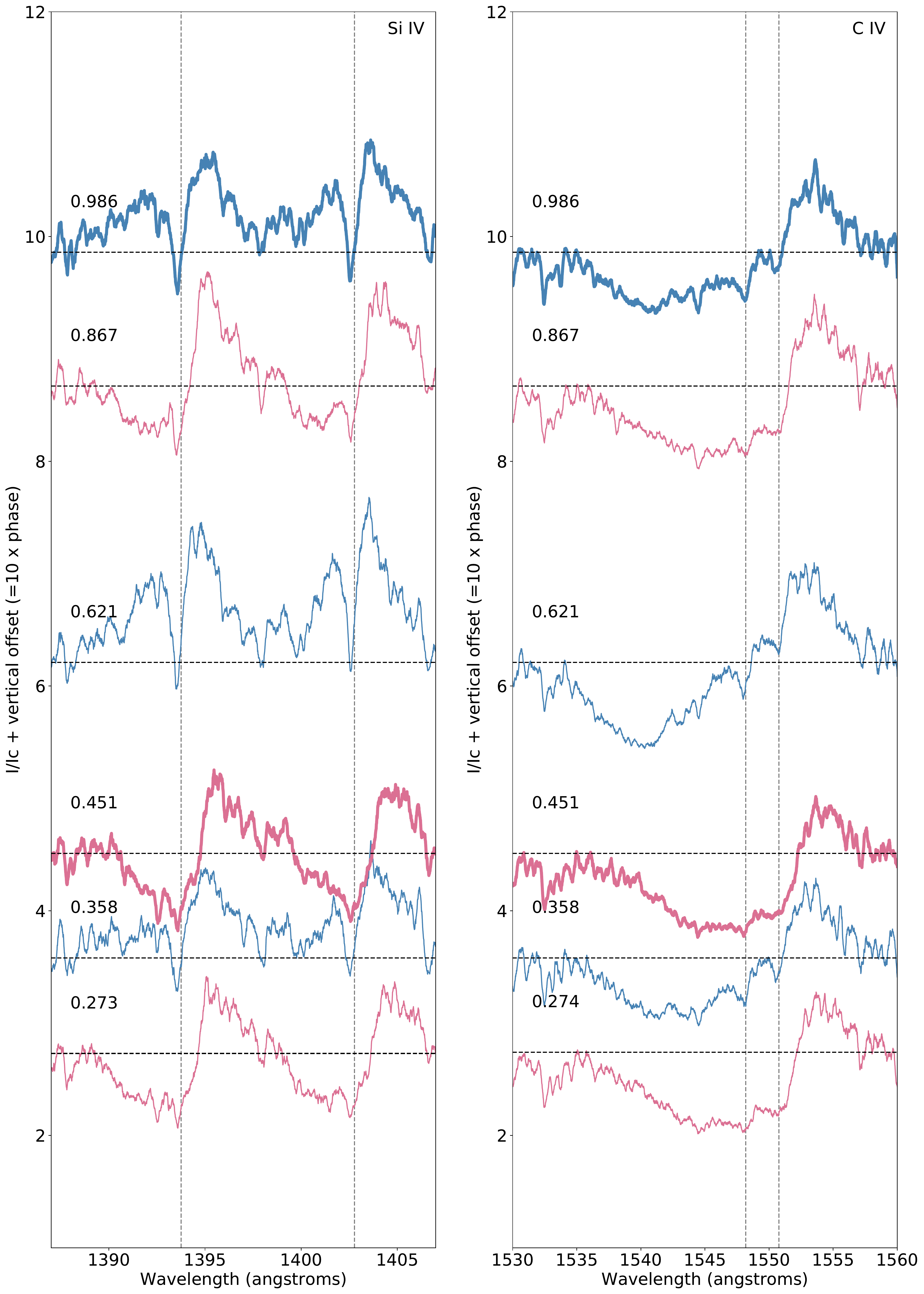}
\caption{Si~\textsc{iv} (left) and C~\textsc{iv} (right) line profiles from HST/COS observations of NGC 1624-2 obtained at six different rotational phases, with a vertical offset proportional to each phase. The phases were computed using the ephemerides of Wade et al. (2012). The thick lines correspond to data published by David-Uraz et al. (2019), while the remaining lines correspond to newly acquired data.
Line profiles with similar morphologies are presented with the same color: blue lines exhibit a ``pole-on"-like appearance, while magenta lines exhibit an ``equator-on"-like appearance.}
\label{fig:d2stack}
\end{figure}

Follow-up UV observations of NGC 1624-2 were obtained with HST (GO program 15066) for four additional rotational phases that sample viewing angles ranging from magnetic pole to equator, yielding an increasingly detailed view of the magnetosphere (see Fig. \ref{fig:d2stack}). The change in the morphology of the line profiles is readily apparent in the C~\textsc{iv} and Si~\textsc{iv} spectral lines. 
These new spectra pose a particular challenge to our understanding of NGC 1624-2's magnetosphere: assuming an ADM-like structure arising from a pure magnetic dipole, one would expect fairly smooth variations of the line profiles across a rotational cycle (as the viewing angle changes). However, we see instead multiple observations obtained at separate phases showing the same line morphology. This hints at a complex magnetospheric structure, which could result from non-trivial surface field configuration; this will be explored in greater detail in a forthcoming publication.


\section{Conclusions and Future Work}


Our initial results show that synthetic line profiles generated using a simplified radiative transfer scheme coupled with the ADM model 
agree well with 
other methods, suggesting a successful new approach capable of accurately reproducing the observed behavior of the wind line profiles of slowly rotating magnetic massive stars without the computational 
cost of using MHD simulations. 
Our synthetic line profiles can help to interpret observational data and provide direct constraints on the properties of massive star magnetospheres. 
Our models also have the potential to be used to identify additional magnetic massive star candidates based on their UV spectra. 
This will prove particularly useful in light of the larger observational surveys such as the UV Legacy Library of Young Stars as Essential Standards (ULLYSES) project, which will observe O- and B-star populations in the (Large and Small) Magellanic Clouds.










\acknowledgements{

CE acknowledges graduate assistant salary support from the Bartol Research Institute in the Department of Physics, University of Delaware, as well as support from programs HST-GO-13629 and HST-GO-15066 that was provided by NASA through a grant from the Space Telescope Science Institute. ADU acknowledges support from the Natural Sciences and Engineering Research Council of Canada (NSERC).

}


\bibliographystyle{ptapap}
\bibliography{cerba}

\end{document}

%% file: shortcuts.tex
\newcommand{\ha}{H$\alpha \;$}

\def\k0{\kappa_{0}}